\documentstyle[12pt]{article}

\begin{document}

%\begin{center}
%{\bf DRAFT}
%\end{center}

\begin{center}
\Large{\bf X-ray Emitting QSOs Ejected from Arp 220}
\end{center}

\begin{center}
{\bf H.C. Arp}\\
Max-Planck-Institut f\"ur Astrophysik, 85741 Garching, Germany
\end{center}

\begin{center}
{\bf E.M. Burbidge} \\
University of California San Diego, Center for Astrophysics and Space Science,
La Jolla CA 92093-0424\\
\end{center}

\begin{center}
{\bf Y. Chu}\\
Center for Astrophysics, University of Science and Technology of China, Hefei,
Anhui 230026, China
\end{center}

\bigskip

\section*{Abstract}

Four compact ROSAT X-ray sources very close to the nearby ultraluminous
infrared galaxy Arp 220 (IC 4553) have been identified as medium-redshift
QSOs. The closest pair lying symmetrically across the galaxy have almost
identical redshifts z = 1.25, 1.26. All of the evidence suggests that these
QSOs have been ejected from Arp 220 and have large intrinsic redshifts.

\medskip

\noindent {\it Subject headings}: Key Words: Galaxies: distances and
redshifts - - - Galaxies: individual(Arp 220) - - - Quasars: general
 - - - X-rays: galaxies

\medskip

\section*{INTRODUCTION}

Arp 220 (IC 4553) is a nearby ultraluminous infrared galaxy (z = 0.018)
that is a strong X-ray source. Close to its center there are a number of
fainter objects with several different, larger, redshifts. Ohyama et al.
(1999) have published optical images and have given spectroscopic data on
the components of this rich field, which they have shown in their Figures
1- 4. One of us, (HCA), has mapped the X-ray field around the galaxy from the
photon event files in the archives of the ROSAT satellite - as shown in
Figures 17, 18 and 19  in Arp (2001a).

Of particular interest are four compact ROSAT X-ray sources which coincide
with stellar images seen on the Palomar Observatory Sky Survey. The
brightest within 50 arc min radius have been named 20.3N and 20.3S (the
numbers corresponding to the ROSAT counts per kilosecond). The brightest in
the inner field are Arp No. 2 and Arp No. 9. Their remarkably exact alignment 
across
Arp 220 is shown in Figure 1. The above four X-ray sources appear to be two
roughly symmetrically placed pairs which we show in Figure 2. The basic
data, 2000 coordinates and magnitudes of the optical objects, are given in
Table 1, together with the coordinates and magnitudes of three faint
stellar objects discovered and discussed by Ohyama et al. (1999).

In Section 2 we give the spectroscopic data for Arp Nos. 2 and 9, and 20.3N
and 20.3S, showing that they are all QSOs. In Section 3 we relate these
results to similar detections associated with other active galaxies.

\section*{ Positions, Magnitudes and Spectroscopy}

Figure 1 shows the ROSAT X-ray data which led to the identification of the
optical objects Arp Nos. 2 and 9, labeled RSO (red stellar object) and BSO
(blue stellar object). They are seen to be  essentially equidistant at 7.0 and 8.0 
arc min from
the catalogued center of Arp 220. In Figure 2 we show a plot of the
locations of these X-ray sources, and also the outer bright pair of ROSAT
sources with exactly equal numbers of counts/ksec, 20.3N and 20.3S.

In Table 2 we give our spectroscopic data for these four QSOs. For Arp Nos.
2 and 9 (RSO and BSO), and 20.3S, observations were made with the Kast
spectrograph on the 3-m Shane Telescope at Lick Observatory, and for 20.3N,
the data were obtained with the OMR spectrograph on the 2.16-m telescope at
Xinlong Station, Beijing Astronomical Observatory. Figure 3 shows the
remarkably similar spectra of Arp Nos. 2 and 9. This similarity is indeed
striking. Our surprise when we observed No. 9, after No. 2, and saw from
the quick-look data the same three broad emissions coming up at the same
wavelengths, can well be imagined.

It is clear from these results that we have found two pairs of QSOs, one
pair of objects at almost exactly the same angular distance from Arp 220
with redshifts that differ only by less than z = 0.009, and a second
pair at distances of 29.8 and 43.1 arc min respectively (Figure 2). As far
as the first pair is concerned, the similarity of the redshifts and the
fact that they lie along an axis through the nucleus of Arp 220 and are at
roughly the same distance in opposite directions makes it highly unlikely
that this is an accidental configuration. The second pair is not as
symmetrically placed, and the redshifts are not equal. Thus, formally the
probability that this configuration is accidental is higher than it is for
the first pair. However, both of the QSOs, 20.3N and 20.3S are bright
(16.3m and 17.7m), and the surface density of QSOs as bright as this is
very low (0.1/square degree for magnitudes 16 and 17 (Kilkenny et al. 1997;
Boyle et al. 1990; Goldschmidt et al. 1992). Thus, even ignoring the
geometrical configurations, to find such bright unrelated QSOs in an area
of $\sim$ 2 square degrees is highly unlikely.

The most likely explanation is that all four of these QSOs are physically
associated with Arp 220, and have been ejected from it, so that the
redshifts are largely intrinsic in origin.

Heckman et al. (1996) have suggested that part of the diffuse X-ray
emission from Arp 220 is due to a faint compact group of galaxies about 2'
southwest of the center. We confirm this as shown here in our Fig 1. Ohyama
et al. (1999) have obtained spectra of these three and have shown that they
are galaxies showing stellar absorption features at a redshift z = 0.09. In
carrying out their investigation they found three faint objects (m = 24.5)
which they call galaxies, but in which the only features that are detected
are emission lines due to Mg II, [O II], [O III], [NeIII] and, in one case, H$\beta$. They
have redshifts z = 0.528, 0.529 and 0.523. The spectra suggest that the objects
can be classified as AGN or starburst nuclei and not normal background
galaxies. They are thus indicated to be related to the other QSOs around Arp 220.
These redshifts are close to that of the QSO 20.3S and lie about 2'.5 from the
center of Arp 220. They are included in Table 1.

\section*{Evidence for Ejection}

By now it has been long accepted that radio jets and lobes, often accompanied
by X-ray jets and X-ray emitting material, are ejected, in generally opposite
directions, from active galaxies. Since compact, energetic X-ray sources must be
relatively short lived it is natural to suppose the excess density of X-ray sources
around active galaxies (Radecke 1997) have been mostly recently ejected. Their
concentration around Arp 220 further supports this conclusion (Arp 2001a).

The outer pair of quasars here confirm the usual pattern of ejected quasars. If
their redshifts of z = .232 and .458 are corrected to the reference frame of Arp 220
(z = .018) the redshifts come out z = .432 and .210. That gives a mean redshift
of z = .32 which is almost exactly the quantized value of z = .30! This enables
one quasar to have been ejected toward us with a line-of-sight
velocity component of v = +.10c and one away from us with v = -.07c. This is
close to what has been found for the 10 previous best pairs discussed by
Narlikar and Arp (2001).

Why then are the two inner quasars so close in redshift at z = 1.25 and 1.26?
An obvious possibility is that they have been ejected across the line of sight. It
will be mentioned  in the next section, however, that there are other close
redshift pairings which strain somewhat the statistical probability of orientation.
A more reasonable expectation is that when ejection does not occur along the
usual path of least resistance, the minor axis direction, that the quasar or proto
quasar interacts strongly with the material of the ejecting galaxy and is slowed
drastically in its ejection velocity. This would furnish a very natural explanation
for the extreme disruption of, in this case, Arp 220. (See also Arp 1999 for a
similar case.) Another advantage of this explanation is that the unusually red
color of QSO No. 2 (O-E = 2.18 mag.) could be naturally explained as entrained
dust from the enormous amounts of dust associated with Arp 220.

Nos. 2 and 9 have redshifts almost exactly between the z = .96 and z = 1.41
peak. That is quite unusual - perhaps they are caught in a short moment of
transit between two major peaks.

Finally the group of strong X-ray galaxies about 2 arcmin SW of Arp 220 are
indicated to be attached by HI radio isophotes (measures by J. Hibbard, see Arp
2001a). They are clearly continuous with Arp 220 in X-rays as shown here in
Fig. 1. At their conventional redshift distance with z = .09 they would be
excessively luminous - close to supposed conventional quasar luminosities! It is
suggested here that they represent ejected material which has been stopped
close to Arp 220. Patches of X-ray material are seen in Fig. 1 to lead from Arp
220 in a somewhat curved track down through this group almost exactly to the
SE QSO, No. 9.

\section*{Discussion and Conclusions}

We have presented new observational evidence showing that two pairs of
X-ray QSOs are associated with the comparatively nearby active galaxy Arp
220, and most likely have been ejected from it.

This is not the first evidence of this kind. Earlier we have shown that a
similar pair of X-ray emitting QSOs lie at approximately the same angular
distance from NGC 4258 and were most likely ejected from it (Burbidge 1995).
A similar but wider pair was also identified across NGC 2639, differing in z by
only 0.018, (Burbidge 1997). Lines of ejected quasars have been reported across
the Seyfert Galaxies NGC 3516 and NGC 5985 (Chu et al 1998; Arp 1998).
Recently three quasars of z = .358, .376 and .380 have been reported around the
X-ray jet galaxy NGC 6217. The latter two quasars with only a difference of .004 in
z are aligned rather well across the galaxy. (See Arp 2001b.) Two of these
quasars are extremely red, suggesting interaction with material in the galaxy as
was inferred for Arp 220.

It is interesting to note that NGC 6217 mentioned above is also catalogued
as Arp 185. This means that the morphological classifications of Peculiar
Galaxies (Arp 1966) between about Arp 140 and Arp 230 are turning up many
highly significant associations with quasars and high redshift companions. Just
the ones which have happened to be investigated so far are: 152, 157, 189,
212,  220 and 227.

%, and a chain of five QSOs that were identified
%initially from their X-ray properties were found apparently ejected from
%the classical Seyfert galaxy NGC 3516 (Chu et al. 1998). Also, there are by
%now many other cases in which the density of QSOs around active galaxies
%strongly suggests that they are physically associated with the galaxy and
%were ejected from it (cf. the case of NGC 1068 (Burbidge 1999), and
%discussion in Arp (1998).

%It appears to us that the inescapable conclusion is that many QSOs are not
%at the distances suggested by their redshifts, but have large intrinsic
%components, and are associated with parent AGN galaxies.

%If we accept this conclusion, but maintain the next most conservative
%assumption, that the parent galaxies lie at the distances obtained by
%supposing that their redshifts are cosmological, we would be looking at
%QSOs with a wide range of absolute magnitudes. For example,

If the distance of NGC 4258 is 7 Mpc and Arp 220 with a redshift z = 0.018 is at a
distance of 90 Mpc (Ho = 60 km/sec Mpc-1) then we would have to accept QSOs
with a wide range of absolute magnitudes. It has been argued, however that Arp
220 is much closer than its redshift distance (Arp 2001a) and it may then be
possible to deal with a much more homogeneous population of low luminosity,
high intrinsic redshift objects.

%Thus, when we find comparatively faint QSOs associated with closeby
%galaxies we are probing a population of intrinsically much fainter QSOs
%than is studied in the usual population. For example, the QSOs which have
%been associated with NGC 4258 would be very faint, and very close to the
%nucleus (in angular measure) if they were coming out of Arp 220. In fact,
%they would be similar to some of the objects found by Ohyama et al. which
%are very close in with redshifts of ~ 0.52. We believe that these objects
%are QSOs or AGN.

\medskip

\centerline{\bf References}
\bigskip

\noindent Arp, Halton 1966, {\it Atlas of Peculiar Galaxies},
California Institute of Technology
\medskip

\noindent Arp, H. C. 1998, {\it Seeing Red: Redshifts, Cosmology and Academic
Science}, Apeiron, Montreal
\medskip

\noindent Arp, H. C. 1999, ApJ 525, 594
\medskip

\noindent Arp, H. C. 2001a, ApJ (1 March, in press)
\medskip

\noindent Arp, H. C. 2001b, ApJ (1 March, in press)
\medskip

\noindent Boyle, B. J., Fong, R., Shanks, T., et al. 1990, MNRAS 243, 1
\medskip

\noindent Burbidge, E. M. 1995, A\&A, 298, L1
\medskip

\noindent Burbidge, E. M. 1997, ApJ, 477, L13
\medskip

\noindent Burbidge, E. M. 1999, ApJ, 511, L9
\medskip

\noindent Chu, Y., Wei, J., Hu, J., Zhu, X. and Arp, H.C. 1998, ApJ 500, 596
\medskip

\noindent Goldschmidt, P., Miller, L., La Franca, F.,
\& Cristiani, S. 1992, MNRAS, 256, 65P
\medskip

\noindent Heckman, T. M., Dahlem, M., Eales, S. A., et al. 1996, ApJ 457, 616
\medskip

\noindent Kilkenny, D., O'Donoghue, D., Koen, C., et al. MNRAS 287, 867
\medskip

\noindent Narlikar, J. V. and Arp, H.C. 2001, ApJ, to be submitted
\medskip

\noindent Ohyama,Y., Taniguchi, Y., Hibbard, J. E., and Vacca, W. D. 1999,
A.J. 117, 2617

\noindent Radecke, H.-D. 1997, A\&A 319, 18
\medskip

\medskip
\pagebreak

\begin{tabular}{|c|c|c|c|c|c|c|c|c|}
\hline
\multicolumn{9}{c}{\bf{TABLE 1}} \\
\hline
\multicolumn{9}{c}{\bf{QSOs and AGN Close to Arp 220}} \\
\hline
Object & \multicolumn{3}{c}{$\alpha$(2000)} &   \multicolumn{3}{c}{$\delta$(2000)} &  $m$ & Distance from
Center\\
\hline
 & h & m & s & $^o$ & $^{\prime}$ &  $^{\prime\prime}$&  & (arc
 min)\\
\hline
20.3N & 15 & 33 & 54.7 & +23 & 56 & 15 & 16.34 & 29.8\\
\hline
Arp 9 & 15 & 34 & 48.1 & +23 & 22 & 31 & 19.82 & 7.0\\
\hline
*Ohyama I & 15 & 34 & 54.3 & +23 & 27 & 59 & 24.5 & 2.33\\
\hline
*Ohyama II & 15 & 34 & 54.5 & +23 & 27 & 52 & 24.3 & 2.43\\
\hline
*Ohyama III & 15 & 34 & 54.8 & +23 & 27 & 44 & 24.7 & 2.54\\
\hline
Arp 2 & 15 & 35 & 6.1 & +23 & 36 & 56 & 19.61 & 8.0\\
\hline
20.3S & 15 & 37 & 14.5 & +23 & 00 & 40 & 17.74 & 43.1\\
\hline
\multicolumn{9}{l}{\bf{*Data from Ohyama et al. (1999).}} \\
\end{tabular}

\medskip\medskip\medskip\medskip

\begin{tabular}{|c|c|c|c|}
\hline
\multicolumn{4}{c}{\bf{TABLE 2}} \\
\hline
\multicolumn{4}{c}{\bf{Redshifts of Four X-Ray QSOs Around Arp 220}} \\
\hline
QSO, 2000 Coordinates &  Line   & $\lambda$(obs) & $z$\\
\hline
Arp No. 9 (BSO) &Mg II 2799   &   6295  &  1.249\\
\hline
 &C III] 1909 &    4290  &  1.247\\
\hline
 &CIV 1549     &   3485   & 1.250\\
\hline
 & &Mean&    1.249\\
\hline\hline
Arp No. 2 (RSO)& Mg II 2799 &     6326   & 1.258\\
\hline
   &     CIII] 1909  &    4311  &  1.258\\
\hline
   &     CIV 1549    &    3496 &   1.257\\
\hline
 & & Mean  &  1.258\\
\hline\hline
20.3 N & H$\alpha$   &   8084  &  0.2318\\
\hline
  &      [O III] 5007  &  6172   & 0.2327\\
\hline
   &     H$\beta$  &   5993  &  0.2329\\
\hline
   & &             Mean  &  0.2325\\
\hline\hline
20.3 S & H$\alpha$   &       9610  &  0.4644\\
\hline
    &    [O III] 5007 &   7320  &  0.4621\\
\hline
    &    [O III] 4959  &  7252 &   0.4624\\
\hline
    &    H$\beta$    &   7108  &  0.4622\\
\hline
     &   H$\gamma$    &   6350  &  0.4631\\
\hline
      &  MgII 2799    &   4093   & 0.4623\\
\hline
    & &            Mean   & 0.4627\\
\hline
\end{tabular}

\section*{Figure Captions}

\noindent Fig. 1 - Hard X-ray band (.5 to 2.4 keV), from ROSAT PSPC, showing
pair of strong sources across Arp 220. Note curved string of sources leading
down to SE QSO. Known redshifts are labeled.

\medskip

\noindent Fig. 2 - Larger field of view in X-rays around Arp 220. In addition to
redshifts some X-ray intensities are labeled in counts/ks. Note strong and equal
count rates for the two outer quasars.
\medskip

\noindent Fig. 3 - Spectra of the two quasars pictured in Fig. 1, z = 1.26 on top
and z =1.25 below. Spectra from 3m Shane telescope of Lick Observatory.

\end{document}